\documentclass[floatfix,showpacs,amsmath,amssymb,letterpaper,groupaddresses,superscriptaddress]{article}
\usepackage{latexsym}
\usepackage{epsfig}

\usepackage{amsmath}
\usepackage{amssymb}
\usepackage{graphicx}
\usepackage{times}
\usepackage[section]{placeins}
\usepackage{caption}
\usepackage{subcaption}
\usepackage{color}

\def\be{\begin{equation}}
\def\ee{\end{equation}}
\def\ba{\begin{eqnarray}}
\def\ea{\end{eqnarray}}
\def\la{\langle}
\def\ra{\rangle}

\def\h{\hskip 1cm}

\def\lo{\longrightarrow}

\begin{document}

\vspace{4cm}
\begin{center}{\Large \bf  Sequential quantum secret sharing in noisy environments}\\
\vspace{2cm}

M. Asoudeh,$^2$\h and \h V . Karimipour.$^1$\\
\vspace{1cm} $^1$ Department of Physics, Sharif University of Technology, P.O. Box 11155-9161, Tehran, Iran.\\
$^2$Department of Physics, Azad University, Northern Branch,  Tehran, Iran .\\

\end{center}

\begin{abstract}
Sequential Quantum Secret Sharing schemes (QSS) do not use entangled states for secret sharing, rather they rely on sequential operations of the players on a single state which is circulated between the players. In order to check the viability of these schemes under imperfect operations and noise in the channels, we consider one such scheme in detail and show that under moderate conditions it is still possible to extract viable secure shared keys in this scheme.  Although we specifically consider only one type of sequential scheme and three different noise models, our method is fairly general to be applied to other QSS schemes and noise models as well.   
\end{abstract}

\section{Introduction}\label{intro}

Quantum Key Distribution (QKD) \cite{BB84, Ekert} and Quantum Secret Sharing (QSS) [3-17]
 are among the most promising areas in the rapidly developing field of quantum technology.   With the growth of demand for secure communication,  it is imaginable that in the near future these two protocols will soon be integrated parts of  modern communication systems.  Both these schemes try to share random sequences of bits between two or more parties in a secure way so that they can use these sequences as a key for encryption and decryption of  messages in  their further communications. In this sense,  QKD is a special case of QSS. In general if there are N players involved in a QSS scheme, the final result is that each player $R_i$, ($i=1, \cdots N$) acquires a random strings of $n-$ bits, denoted by $K_i$ so that 
\be
K_1+K_2+\cdots K_N=0
\ee
where the summation is bitwise and modulo 2. Therefore when a dealer say $R_1$ wants to send a message $M$ to the other players $R_2, \cdots R_N$, so that they can only retrieve the message by their full collaboration, he encrypts the message as
\be
M\lo M+K_1
\ee
and sends it to the players who will add their own $K_i$ to make it non-retrievable for any subset of players and retrievable only to the last (authorized player $R_N$) who, finds the message in the form

\be
M+K_1+K_2+\cdots K_N=M.
\ee

The main ingredient of almost all the QSS protocols is a multi-partite entangled state, like a Greenberger-Horne-Zeilinger (GHZ) state which acts as splitter of information. In view of the extreme
fragility of these states and the difficulty for their preparation, there has been a series of attempts to devise QSS protocols which do not need any entanglement \cite{Schmidt,YanGao,He, KA, Tavak}. \\

While there has been some studies on the performance of entanglement-based
QSS schemes \cite{chen, adesso}, it seems that there has been almost no studies of the more promising and more practical sequential QSS schemes. There is a  report on the effect of noise on a  sort of QSS scheme \cite{chat}, however it is not really related to the protocol that we discuss here, since the dealers manipulate entangled states.   A related report is that of \cite{agrawal} who studies a protocol with three parties. We should stress that in the protocol \cite{Schmidt} that we study here, only a single qubit is circulated between the parties with no entanglement at any stage of the protocol.  It is important to note that in view of the highly fragile character of GHZ and other entangled states \cite{bohmann}, it seems that if QSS schemes become practical in the future, they will be of the sequential rather than the entanglement-based type. Therefore it is highly desirable to see how in a noisy environment, errors accumulate and how the final error depends on the number of players. \\

 Here we study noise effects in QSS scheme of \cite{Schmidt} for arbitrary number of players. We consider two classes of noise. The first class concerns imperfections in the unitary actions of the players and  the second class concerns the noise in the channels between consecutive players. In this later class we study the effect of a number of well-known and physically motivated channels, namely de-phasing, depolarizing, phase-flip and bit-flip channels. We find that  in both classes of imperfections  and channel noises, the required precision for establishing a reliable secret scales inversely with the number of players. \\

The structure of the paper is as follows. In section \ref{seq} we briefly describe the general idea of sequential QSS \cite{Schmidt,KA,Tavak} with emphasis on the special case of the scheme of \cite{Schmidt}, which will be our model of choice for studying noise effects. To study the sequence of actions of players followed by the effect of noise, we resort to the vectorized formalism for describing states and operations in section \ref{vect}. Then in section \ref{gate}, we briefly study the effect of imperfections of unitary actions of players and then in section \ref{noise}  we use this formalism to study the effect of de-phasing, depolarizing and bit-flip channels respectively. We end the paper with a conclusion. \\

\section{Sequential quantum secret sharing }\label{seq}
Sequential Quantum Secret Sharing (QSS) scheme, tries to avoid the use of multipartite entangled states and instead tries to develop a scheme which is in the same spirit of the BB84 protocol \cite{BB84} for quantum key distribution. That is, it is based on circulating and manipulating specific states among the players which if measured in the right basis by the last player will establish a shared random secret key between the players. More concretely, in such schemes there are $N$ players which we denote by $R_1, R_2, \ $ to $R_N$. The first player prepares a reference state $|\phi_1\ra$ (a specific state in a basis $B=\{|e_0\ra,  |e_1\ra, \cdots |e_{d-1}\ra\}$) in a $d$ dimensional space and acts on it by a unitary operator $U({\bf s}_1, {\bf p}_1)$, where ${\bf s}$ and ${\bf p}$ stand for a number of secret (private) and public parameters. He then passes the state $U({\bf s}_1, {\bf p}_1)|e_0\ra$ to player $R_2$ who acts similarly until after a full circle the state comes back to the first player in the form 

\be
|\psi_f(\{{\bf s}, {\bf p}\})\ra=U({\bf s}_N, {\bf p}_N)\cdots U({\bf s}_2, {\bf p}_2)U({\bf s}_1, {\bf p}_1)|e_0\ra
\ee
The players then announce the public parameters ${\bf p}_1$ to ${\bf p}_N$. In a certain fraction of rounds (depending on the scheme, $\frac{1}{2}$ for \cite{Schmidt} and \cite{KA}, and $\frac{1}{d}$ for \cite{Tavak}), which we call valid rounds, the public parameters are such that  the state $|\psi_f(\{{\bf s}, {\bf p}\})\ra$ is one of the states in the basis $B$, say $|e_m\ra$.  Since the first players always measures his state in this basis, in such rounds there will be a precise relation between the secret parameters ${\bf s}_i$ and the final measured parameter $m$, in the form 
\be
F(m, {\bf s}_1, {\bf s}_2, \cdots {\bf s}_N)=0.
\ee
This relation establishes a shared key between the secret parameters of all the players, in which the parameter $m$ is also included. \\ 

The first QSS scheme of this kind was introduced in \cite{Schmidt}, where the players use qubits and their unitary action is a phase gate of the form $U(\phi)|m\ra=e^{im\phi}|m\ra$, where
$\phi\in\{0,\frac{\pi}{2},\phi,\frac{3\pi}{2}\}$, figure (\ref{QSSFigureLabel}). This set of phase gates can be divided into two classes, Class 0=$\{0,\pi\}$ and Class 1=$\{\frac{\pi}{2},\frac{3\pi}{2}\}$. The 1-bit information as to which class has been chosen by a particular player $R_n$ is a public parameter which is denoted by $p_n=0, 1$ and the  information as to which particular phase has been chosen by the player $R_n$ within a class is a secret parameter kept only with $R_n$ to be used at the end of the protocol. Starting from $|+\ra=\frac{1}{\sqrt{2}}(|0\ra+|1\ra)$, the final state will be 
\be\label{Fin}
|\psi_f\ra=\frac{1}{\sqrt{2}}(|0\ra+e^{i(\phi_1+\phi_2+\cdots \phi_N)}|1\ra).
\ee

\begin{figure}[h]\centering
\includegraphics[width=10cm,height=8cm,angle=0]{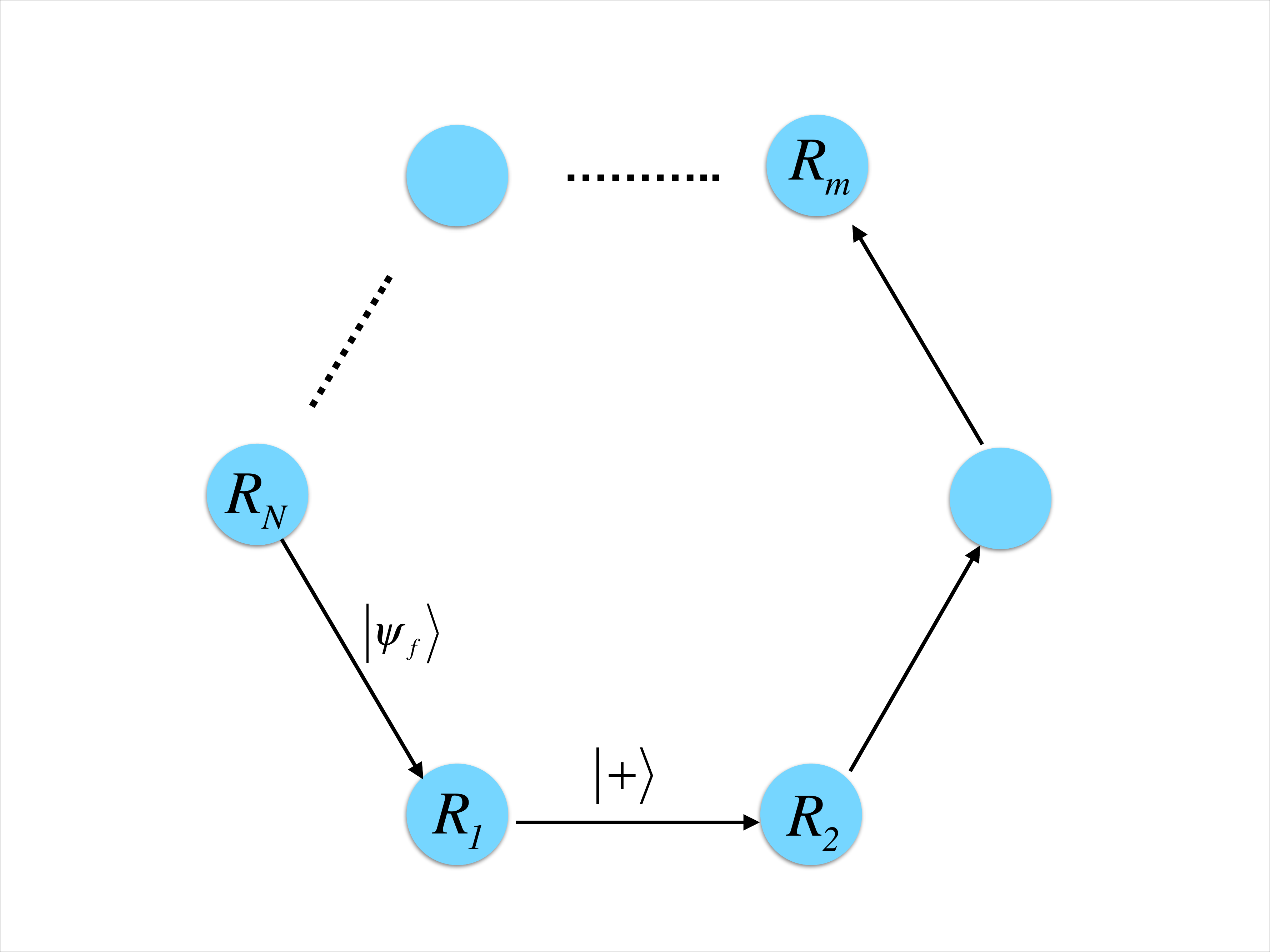}
\caption{The idea case of the sequential QSS scheme of \cite{Schmidt}, where states are acted by perfect gates by the players and are transmitted through noiseless channels.  }
\end{figure}\label{QSSFigureLabel}

At the end of the protocol, all the players announce their class values, $p_i$.  The player $R_1$ or the dealer, measures the final state () in the $X$ basis. When $\sum_i p_i=1\  (mod\  2)$ the final state (\ref{Fin}) is an eigenstate of the $Y$ operator, leading to a probabilistic result (with no correlation) and the round should be discarded as an invalid round. However when $\sum_i p_i=0 (mod \ 2)$, the final state is an eigenstate of the $X$ operator and a perfect correlation exists between the secure parameters and the measurement of the final state. That is 
\be\label{Rem}
\phi_1+\phi_2+\cdots + \phi_N=m,
\ee
where $m$ is the result of measurement of the player $R_1$ in the X basis.  Let us denote the final result of measurement by $m$ ($m=0$, for $+$ and $m=1$ for $-$). Then equations (\ref{Rem}) show that  the shared secret, defined as 

\ba
K_1&=& \frac{\phi_1}{\pi}+m,\cr
K_{i}&=&\frac{\phi_i}{\pi}, \ \ \ i=2, \cdots N,
\ea
will have perfect correlation as follows:  
\be
K_1+K_2+\cdots K_N=0.
\ee

This will then allow the players to share a secure key among the players. Since the work of \cite{Schmidt} many other schemes have also been proposed for sequential QSS \cite{YanGao, KA, Tavak}, where the use of entanglement is bypassed. \\

All this has been described under ideal situation, when the gates act perfectly and the channels are noiseless. In a realistic situations, both assumptions need to be replaced with modest assumptions. Here we assume that the classical channels used for public announcements of public parameters are noiseless, but the gates used by players have some imperfections and also the quantum channels used for transmitting the states are noisy. We want to see how these two kinds of imperfections affect the reliability of this QSS scheme.  

\section{Vectorization of states and operations}\label{vect}
In a sequential QSS scheme, $N$ players are acting on a reference signal one after the other and in each transmission between any two consecutive players, the signal is also affected by noise in the channel. To determine the admissible level of noise in these channels (which at first are assumed to be identical), we have to find the cumulative effect of both the noise and the actions of different players on the initial state. 
The basic question we are faced with is to determine in an analytic form the cumulative effect of noise and the actions of all individual players. As we will see,  the best approach for doing this is to use the vectorized form of quantum channels which for simplicity is described here for qubit states and channels. \\

Let $\rho=\left(\begin{array}{cc} a & b \\ c & d\end{array}\right) $ be a general one-qubit state.  To this corresponds a vector 
\be
|\rho\ra=\left(\begin{array}{c} a \\ b\\ c \\ d\end{array}\right).
\ee
A state like $\rho_0=|+\ra\la +|$ where $|+\ra=\frac{1}{\sqrt{2}}(|0\ra+|1\ra)$ is depicted as 
\be\label{vectin}
|\rho_0\ra=\frac{1}{2}\left(\begin{array}{c} 1 \\ 1 \\ 1 \\ 1\end{array}\right).
\ee

More generally 
a state $\rho=\sum_{i,j}\rho_{i,j}|i\ra\la j|$ can be cast into the vector form $|\rho\ra=\sum_{i,j}\rho_{i,j}|i,j\ra$. The inner product of two matrices $A$ and $B$ is the same as the inner product of their vectorized forms, that is
\be\label{inn}
tr(A^\dagger B)=\la A|B\ra.
\ee

A quantum channel acts  on the state $\rho$ as follows
\be
\rho\lo \rho'=E(\rho)=\sum_k A_k\rho A_k^\dagger,
\ee
where 
\be
\sum_k A_k^\dagger A_k=I.
\ee

This operation can be depicted as  linear map ${\cal E}$ on the vector $|\rho\ra$. In fact 

\be
|\rho\ra\lo |\rho'\ra={\cal E}|\rho\ra,\h {\cal E}=\sum_{k}A_k\otimes A_k^{*}.
\ee
The unitary action $U(s_m,p_m):|\phi\ra\lo |\phi'\ra$ by the $m-$th player, can also be written, first as a quantum channel $\rho\lo U_m(s,p)\rho U_m^\dagger$ and then vectorized in the form

\be
|\rho\ra\lo {\cal U}(s_m,p_m)|\rho\ra= U(s_m,p_m)\otimes U(s_m, p_m)^*|\rho\ra. 
\ee
Combining all these actions we find
\be
|\rho_f\ra=\prod_{m=1}^N ({\cal E}{\cal U}({s_m,p_m}))|\rho_0\ra
\ee
where $|\rho_0\ra$ is the vectorization of the first pure state started by the first player $R_1$ and $|\rho_f\ra$ is the vectorized form of the final state received by $R_1$. 
One can then return back the final state from the vectorized form $|\rho_f\ra$ to the standard matrix form $\rho_f$ and analyze its various properties, although this is not really necessary since all the measurement results and probabilities can also be expressed in vector form as in (\ref{inn}). 

\section{Gate imperfections}\label{gate}
Let us assume that the phases applied by the players are not exactly in the form $\{0,\frac{\pi}{2}, \frac{3\pi}{2}, \pi\}$, but have error with a fixed average. For example suppose that each player $R_i$, instead of the phase $\phi_i$
applies a phase $\phi_i+\epsilon_i$, where $\epsilon_i$ is a random phase with average $\overline{ \epsilon}$. In this case the final state will be 
\be
|\psi_f\ra=\frac{1}{\sqrt{2}}(|0\ra+e^{i\sum_i (\phi_i+ \epsilon_i)}|1\ra).
\ee
We assume that the errors $\epsilon_i$ are not large enough to change the class of the phases and they only randomly shift the phase within a class. Therefore we only have to consider valid rounds where $\sum_i \phi_i$ is supposed to be $0$ or $\pi$ mod $2\pi$. The error is committed when $\sum_i\phi_i=0$ and the measured state is $|-\ra$ or when $\sum_i\phi_i=\pi$ and the measured state is $|+\ra$.  Assuming that in half of the cases $\sum_i\phi_i=0$ and in the other half $\sum_i \phi_i=\pi$, we find for a specific set of $\epsilon_i$'s 

\ba
P_{error}&=&\frac{1}{2}P(-|\sum_i\phi_i=0)+\frac{1}{2}P(+|\sum_i\phi_i=\pi)\cr
&=&\frac{1}{2}\mid \la -|\psi_f(\sum_i\phi_i=0)\ra\mid^2 + \frac{1}{2}\mid \la +|\psi_f(\sum_i\phi_i=\pi)\ra\mid^2 =\frac{1}{2}(1-\cos \sum_i \epsilon_i).
\ea
The average rater of error will be given by 

\be
\overline{P_{error}}=\frac{1}{2}(1-\cos N\overline{\epsilon}).
\ee
As long as $\overline{P_{error}}<< \frac{1}{2}$, one can use a privacy amplification algorithm to extract a shorter error-free key from a long key. In this paper we do not consider these algorithms and only note that  amount of the  shortening of the key depends on the error probability.   Let us demand that $\overline{P_{error}}<\frac{\delta}{2}$, where $\delta$ is a  parameter less than one. Then this means that the phase error is bounded as 

\be
\overline{\epsilon}<\frac{\arccos(1-\delta)}{N}\approx \frac{\sqrt{2\delta}}{N}.
\ee

\section{Effect of noise in the channels}\label{noise}

We now turn to the noise in channels. For simplicity at first we assume that the noise parameters  are identical, but later this assumption will be relaxed. We consider four different physically motivated noise channels. 

\subsection{Phase damping channel}
The first channel that we study is the phase dampling channel
\be
E_1(\rho)=\ (1-p)\rho + p A_0\rho A_0^\dagger + p A_1\rho A_1^\dagger,
\ee
with $A_0=\left(\begin{array}{cc} 1 & 0 \\ 0 & 0\end{array}\right)$ and  $A_1=\left(\begin{array}{cc} 0 & 0 \\ 0 & 1\end{array}\right)$. Such a channel which decreases the phase coherence of the state,  is perhaps the most relevant type of noise both in the present context and in view of practical considerations. It acts on any single-qubit density matrix as follows

\be
\left(\begin{array}{cc} a & b \\ c & d\end{array}\right)\lo \left(\begin{array}{cc} a & (1-p)b \\ (1-p)c & d\end{array}\right).
\ee

\be
{\cal E}_1=\left(\begin{array}{cccc} 1 & & & \\ & 1-p & & \\ & & 1-p&\\ & & & 1 \end{array}\right)
\ee
The action of phase gate can also be depicted by a diagonal matrix ${\cal R}(\phi_i)$

\be
{\cal U}(\phi_i)=\left(\begin{array}{cccc} 1 & & & \\ & e^{i\phi_i} & & \\ & & e^{i\phi_i}&\\ & & & 1 \end{array}\right).
\ee
In view of the vectorized form of the initial state (\ref{vectin}), we find  
\be
|\rho_f\ra={\cal E}{\cal U}(\phi_N)\cdots {\cal E}{\cal U}(\phi_2) {\cal E}{\cal U}(\phi_1)|{\cal \rho}_0\ra=\frac{1}{2}\left(\begin{array}{c} 1 \\ (1-p)^N e^{\sum_m i\phi_m}\\  (1-p)^N e^{\sum_m i\phi_m}\\ 1\end{array}\right)
\ee

In matrix form the final state will be given by
\be
\rho_f=\left(\begin{array}{cc} 1 & (1-p)^N e^{\sum_m i\phi_m}\\ (1-p)^N e^{\sum_m i\phi_m}& 1\end{array}\right)
\ee

This shows that when, after a full round, the first player measures the state in the $X$-basis, there is not perfect correlation anymore. In fact the first player obtains both $|+\ra$ and $|-\ra$ with the following conditional probabilities:

\ba
P(+|\sum_i \phi_i)&=&\la +|\rho_f|+\ra=\frac{1}{2}\left(1+(1-p)^N\cos (\sum_{i=1}^N\phi_i)\right)\cr
P(-|\sum_i\phi_i) &=&\la -|\rho_f|-\ra=\frac{1}{2}\left(1-(1-p)^N\cos (\sum_{i=1}^N\phi_i)\right).
\ea
 It can be seen that in the absence of noise (when $p=0$), there is perfect correlation between the result of the last measurement and the sum of phases. For example we have 
 $
P(+\mid \sum_i \phi_i=0)=1, \  $ and $\ \   P(-\mid \sum_i\phi_i=0)=0.
$

The presence of noise decreases this correlation and leads to errors. Assuming that all the phases are chosen randomly by the players, we have $P(\sum_i \phi_i)=\frac{1}{2}$ and hence the probability of error is  given by
\be\label{perror1}
P_{error}=\frac{1}{2}P(-\mid  \sum_i \phi_i=0)+\frac{1}{2}P(+\mid  \sum_i \phi_i=\pi)=\frac{1}{2}(1-(1-p)^N)
\ee
Again if we demand that $P_{error}<\frac{\delta}{2}$, this leads to the following bound for the noise parameter $p$, 

\be\label{bound1}
p< 1- (1-\delta)^{\frac{1}{N}}\approx \frac{\delta}{N},
\ee
showing that the level of admissible noise decreases inversely with the number of players. \\

If the noise parameters are not equal which is quite expected in view of different distances between the players, then a simple look at the previous analysis leads to generalization of (\ref{perror1}) with the following 

\be
P_{error}=\frac{1}{2}\left(1-\prod_{i=1}^N(1-p_i)\right).
\ee
This shows that if one of the channels has a large noise, i.e. $p_i\approx 1$, then $P_{error}\approx \frac{1}{2}$, rendering the whole scheme useless. To understand this effect note that assume that the $i-$th channel has $p_i=1$. Then any previous state which is necessarily in the form $\frac{1}{2}\left(\begin{array}{cc} 1 & c\\ c^* & 1\end{array}\right)$, when passing through this channel becomes equal to $\frac{1}{2}I$ and this completely mixed state is never affected by the actions of the subsequent players, leading to no correlation at the end of the protocol.

\subsection{Depolarizing channel}
The second channel that we consider is the depolarizing channel defined as 

\be
E_2(\rho)= (1-p)\rho + \frac{p}{2} tr(\rho) I .
\ee
To vectorize this channel we note that for a qubit $ tr(\rho)=a+d$, and rewrite this as 
\be
\left(\begin{array}{cc} a & b \\ c & d\end{array}\right)\lo \left(\begin{array}{cc} (1-\frac{p}{2})a+\frac{p}{2}d & (1-p)b \\ (1-p)c & (1-\frac{p}{2})d+\frac{p}{2}a\end{array}\right).
\ee
which shows that the vectorized form of the channel is given by
\be
{\cal E}_2=\left(\begin{array}{cccc} 1-\frac{p}{2} & && \frac{p}{2} \\ & 1-p & & \\ & & 1-p& \\ \frac{p}{2} && & 1-\frac{p}{2}\end{array}\right).
\ee 
The combination of this channel with the unitary action of each player is given by
\be
{\cal E}_2 {\cal U}(\phi_m) = \left(\begin{array}{cccc} 1-\frac{p}{2} & && \frac{p}{2} \\ & (1-p)e^{i\phi_m} & & \\ & & (1-p)e^{-i\phi_m}& \\ \frac{p}{2} && & 1-\frac{p}{2}\end{array}\right).
\ee 
The product of all these matrices is then given by

\be
\prod_{m=1}^N {\cal E}_2 {\cal R}(\phi_m)= \left(\begin{array}{cccc} \frac{1+\eta^N}{2} & & & \frac{1-\eta^N}{2}\\ & (1-p)^N e^{i\sum_m \phi_m} & & \\ & & (1-p)^N e^{-i\sum_m \phi_m}&\\ \frac{1-\eta^N}{2}& & & \frac{1+\eta^N}{2} \end{array}\right)
\ee

where
$
\eta=(1-p).
$
The final density matrix $|\rho_f\ra$ is obtained by acting this linear operator on the vectorized form of the state $|+\ra\la +|$ . This state which will be measured by the player $R_1$ is given by 
\be
|\rho_f\ra=\frac{1}{2}\left(\begin{array}{c} 1 \\ (1-p)^N e^{i\sum_m \phi_m}\\ (1-p)^N e^{-i\sum_m \phi_m}\\ 1 \end{array}\right).
\ee
Although the channels act differently on general input states, it is seen that the final state is the same as the one in the previous subsection. Therefore the same analysis and the same bound is also valid here.

\subsection{Bit flip channel}
The last  channel that we study is the bit flip channel which requires a more detailed analysis. It is defined by
\be
E_3(\rho)= (1-p)\rho + p X\rho X.
\ee
which transforms the state in the following way
\be
\left(\begin{array}{cc} a & b \\ c & d\end{array}\right)\lo \left(\begin{array}{cc} (1-p)a+pd & (1-p)b+pc \\ (1-p)c+pb & (1-p)d+pa\end{array}\right).
\ee
From this we can extract the matrix form of the CPT map $E_2$:
\be
{\cal E}_3=\left(\begin{array}{cccc} 1-p & & & p\\ & 1-p & p& \\ &p & 1-p&\\ p& & & 1-p \end{array}\right)
\ee
The concatenation of the action of the $m-$th player and the action of the channel $E_2$ is given by 
\be
{\cal E}_3 R(\phi_m) =\left(\begin{array}{cccc} 1-p & & & p\\ & (1-p)e^{i\phi_m} & pe^{-\phi_m}& \\ &pe^{i\phi_m} & (1-p)e^{-i\phi_m}&\\ p& & & 1-p \end{array}\right).
\ee
The above matrix has a block form, written as 
\be
{\cal E}_3{\cal U}(\phi_m)=A_m\oplus B_m,
\ee
where 
\be A_m=\left(\begin{array}{cc} 1-p & p \\ p & 1-p\end{array}\right) \ee is the outer block matrix and \be B_m=\left(\begin{array}{cc} (1-p)\ e^{i\phi_m} & p\ \ e^{-i\phi_m} \\ p\ e^{i\phi_m} & (1-p)\ e^{-i\phi_m}\end{array}\right)\ee
is the inner block matrix. This form easily allows us to multiply a sequence of such matrices in a straightforward way.
We readily find
\be
\prod A_m=\frac{1}{2}\left(\begin{array}{cc} 1+\gamma^N & 1-\gamma^N\\ 1-\gamma^N & 1+\gamma^N  \end{array}\right),
\ee
where
\be
\gamma=(1-2p).
\ee

The calculation of $\prod_m B_m$ is a little tricky. First we note that 
\be
B_m=\left((1-p)I+pX\right)e^{i\phi_m Z}.
\ee
In general (i.e. for arbitrary values of $\phi_m$)  it is  difficult to simplify a product of such matrices, since the passing of $e^{i]\phi_m Z}$ creates complicated terms when comulated after many passings. However we note that $\phi_m$ takes only specific values from the set \be\{0,\frac{\pi}{2}, \pi, \frac{3\pi}{2}\}\equiv  \{I, iZ, -I, -iZ
\}.
\ee 
This can be written compactly as $$e^{i\phi_m Z}=(-1)^{s_m}(iZ)^{p_m},$$ where $s_m\in \{0,1\}$ and $p_m \in \{0,1\}$ are respectively the secret and public parameters of the player $R_m$. In fact the parameter $p$ shows the class and $s$ shows the phase within the class.  
$p$ is made public and $s$ is kept secret. 
 In order to calculate $\prod_m B_m$, we rewrite $B_m$ as follows:
\be
B_m=\left((1-p)I+pX\right)(-1)^{s_m}(iZ)^{p_m}.
\ee

It is now easy to pass $Z^{p_m}$ through different terms in the product. For example we find 

\ba
B_1\ B_2&=&(-1)^{s_1+s_2}\left((1-p)I+pX\right)(iZ)^{p_1}\left((1-p)I+pX\right)(iZ)^{p_2}\cr
&=&(-1)^{s_1+s_2}\left((1-p)I+pX\right)\left((1-p)I+p(-1)^{p_1}X\right)(iZ)^{p_1+p_2}
\ea
Continuing in this way, we find:
\be
\prod_m B_m =(-1)^{s_1+s_2+\cdots s_N}\prod_{m=1}^N\left[(1-p)I+p\xi_{m-1} X\right] (iZ)^{\sum_{m=1}^N p_m}
\ee
where
\be
\xi_m=(-1)^{{\sum_{i=1}^{m}p_{i}}}.
\ee
Hereafter we  consider only the case of valid rounds where $\sum_{m=1}^N p_m=0$ which considerably simplifies subsequent calculations. 
Using the fact that $HXH=Z$, where $H=\frac{1}{\sqrt{2}}\left(\begin{array}{cc} 1 & 1 \\ 1 & -1\end{array}\right)$ is the Hadamard matrix, we find

\be
\prod_m B_m =H\left(\begin{array}{cc} a_N &\\ &  b_N \end{array}\right)H
\ee
where
\be\label{ab}
a_N=\eta_N\prod_{m=1}^N(1-p+p\xi_{m-1})\ \ \ , \ \ \ b_N=\eta_N\prod_{m=1}^N(1-p-p\xi_{m-1}).
\ee
in which we have used $\sqrt{\xi_N}$ as a short way of writing $(i)^{\sum_{m=1}^N p_m}$ and $\eta_N$ is determined by the sum of secret parameters: 

\be
\eta_N:=(-1)^{\sum_{m=1}^N s_m}.
\ee

Multiplying the Hadamard matrices, the product of inner matrix blocks is found: 
\be
\prod_m B_m =\frac{1}{2}\left(\begin{array}{cc} a_N +b_N& a_N-b_N \\ a_N-b_N& a_N+b_N\end{array}\right).
\ee

The final density matrix in the vectorized form is now given by
\be
|\rho_f\ra=\frac{1}{2}\left(\begin{array}{cccc}1+\gamma^N && & 1-\gamma^N\\ & a_N+b_N & (a_N-b_N)& \\ & a_N-b_N & (a_N+b_N)& \\ (1-\gamma^N) && & (1+\gamma^N)\end{array}\right)\frac{1}{2}\left(\begin{array}{c} 1 \\ 1 \\ 1 \\ 1\end{array}\right)
\ee
 or
 \be
 |\rho_f\ra=\frac{1}{2}\left(\begin{array}{c}1 \\ a_N \\ a_N\\ 1\end{array}\right)
 \ee
 In matrix form this density matrix is given by:
 \be
 \rho_f=\frac{1}{2}\left(\begin{array}{cc} 1 & a_N\\ a_N & 1\end{array}\right).
 \ee

We can now calculate the error probability. To this end we note that contrary to the previous cases, the probability of results of measurement in the $X$ basis depend, not on the sum of parameters but on each of them. This is due to the relation (\ref{ab}). So we find 

\ba
P(+|\{s_i\})&=&\frac{1}{2}(1+a_N)=\frac{1}{2}\left(1+(-1)^{\sum_{m=1}^N s_m}\prod_{m=1}^N(1-p+p\xi_{m-1})\right)\cr
P(-|\{s_i\})&=&\frac{1}{2}(1-a_N)=\frac{1}{2}\left(1-(-1)^{\sum_{m=1}^N s_m}\prod_{m=1}^N(1-p+p\xi_{m-1})\right).
\ea
As a test we note that when $p=0$, we find that

\ba
P(+|\{s_i\})&=&\frac{1}{2}(1+(-1)^{\sum_{m=1}^N s_m})=\delta(\sum_{m=1}^N s_m,0),\cr
P(-|\{s_i\})&=&\frac{1}{2}(1-(-1)^{\sum_{m=1}^N s_m})=\delta(\sum_{m=1}^N s_m,1),
\ea
which shows that when $\sum_{m=1}^N s_m=0$, the measurement result is definitely $+$ and when 
$\sum_{m=1}^N s_m=1$, then the measurement result is definitely $-$. This is the perfect correlation which is expected in the absence of noise. \\

  Let us denote by $\{s_i\}_0$, the set of all parameters  $\{s_1, s_2, \cdots s_N\}$ such that their sum equals zero, i.e. $\{s_i\}_0=\{s_1, s_2, \cdots s_N\ \mid\  s_1+s_2+\cdots s_N=0 \}$ with a similar definition for $\{s_i\}_1$. Then the conditional probabilities that we need are the following:  

\ba
P(+|\{s_i\}_1)&=&\frac{1}{2}\left(1-\prod_{m=1}^N(1-p+p\xi_{m-1})\right)\cr
P(-|\{s_i\}_0)&=&\frac{1}{2}\left(1-\prod_{m=1}^N(1-p+p\xi_{m-1})\right).
\ea
The equality of these two terms makes the calculation of the final error probability feasible, since, using 
$\sum_{\{s_i\}_0}+\sum_{\{s_i\}_1}=\sum_{s_1,s_2,\cdots s_N}$,
 we can write
\ba
P_{error}&=&\sum_{\{s_i\}_0}P(s_1,s_2,\cdots s_N)P(-|\{s_i\}_0)+\sum_{\{s_i\}_1}P(s_1,s_2,\cdots s_N)P(+|\{s_i\}_1)\cr
&=&\frac{1}{2}\sum_{s_1,s_2,\cdots s_N}P(s_1,s_2,\cdots s_N) \left(1-\prod_{m=1}^N(1-p+p\xi_{m-1})\right)
\ea

Naturally, due to independence of the actions of players, we have the uniform  distribution $P(s_1,s_2, \cdots s_N)=\frac{1}{2^N}$. We also note that all the parameters $\xi_m$ are also independent, taking the values $1$ and $-1$ with equal probabilities. Therefore (in view of $\xi_0=1$) we find

\be
\sum_s(1-p+p\xi_0)=\sum_s(1-p+p)=2\ \ \ \ 
\sum_s(1-p+p\xi_m)=\sum_s(1-p+p\xi_m)=2-2p.
\ee
Putting everything together, the final error probability will be 
\be\label{perrorb}
P_{error}=\frac{1}{2}\left(1-(1-p)^{N-1}\right).
\ee
This leads to the same bound as in (\ref{bound1}) with $N$ replaced with $N-1$.\\

In view of the lengthy calculations of this case, let us study the simplest case in a concrete way, where $N=2$ and follow the operations from the beginning. The player $R_1$ prepares the state $|\psi_0\ra=|+\ra$ and acts on it by his phase gate to produce $|\psi_1\ra=\frac{1}{\sqrt{2}}(|0\ra+e^{i\phi_1})|1\ra$. This state undergoes the bit flip noise and becomes 
\be
\rho_1 = \frac{1}{2}\left(\begin{array}{cc} 1 & (1-p)e^{-i\phi_1}+pe^{i\phi_1}\\ (1-p)e^{i\phi_1}+pe^{-i\phi_1}&1\end{array}\right).
\ee
When reached to $R_2$ it is acted by a second phase gate $R(\phi_2)$ and becomes
\be
\rho_1 = \frac{1}{2}\left(\begin{array}{cc} 1 & (1-p)e^{-i\phi_1-i\phi_2)}+pe^{i\phi_1-i\phi_2}\\ (1-p)e^{i\phi_1+i\phi_2}+pe^{-i\phi_1+i\phi_2}&1\end{array}\right)=:\frac{1}{2}\left(\begin{array}{cc} 1 & a \\ \overline{a}& 1\end{array}\right)
\ee
where the last equality defines the parameter $a$.
Before deriving the final matrix, when noise acts on this matrix, let us find the density matrix at this stage for various values of $\phi_1$ and $\phi_2$ in the valid rounds. These values are shown in table  (\ref{QT}) together with the form of the matrix $\rho_1$. It is now clearly seen that all of these later states are invariant under bit-flip noise and so the final density matrix $\rho_f$ is the same as $\rho_1$ shown in table (\ref{QT}). The last three columns show probabilities of obtaining $+$ and $-$ in the final measurement of $X$ by the first player $R_1$  and the probability of error. Note that in each row, the superscript $*$ denotes which result is expected (in view of the parameter $\phi_1+\phi_2$). This easily  determines the error probability in each case:

	\begin{table}
	
		\centering
	\begin{tabular}{||c c c c c c c||} 
		\hline
		$\phi_1$ & $\phi_2$ & $a$ & $\rho_1$ & $P(+)$ & $P(-)$ &  $P_{error}$\\ 
		\hline\hline
		$0$ & $0$ & $ 1$& $|+\ra\la +|$ & $1^*$ & $0$ &  $0$\\ 
		\hline
		$0$ & $\pi$ & $ -1$& $|-\ra\la -|$ &$0$ & $1^*$ &  $0$\\
		\hline
		$\pi$ & $0$ & $ -1$& $|-\ra\la -|$ & $0$ & $1^*$ &  $0$\\
		\hline
		$\pi$ & $\pi$ & $1$&$|+\ra\la +|$ & $1^*$ & $0$ &  $0$\\ 
		\hline
		$\frac{\pi}{2}$ &$\frac{\pi}{2}$ & $2p-1$ &$\frac{1}{2}\left(\begin{array}{cc} 1 & 2p-1 \\ 2p-1& 1\end{array}\right)$ & $p$ & $(1-p)^*$ &  $p$\\
		\hline
		$\frac{\pi}{2}$ &$\frac{3\pi}{2}$ & $1-2p$ &$\frac{1}{2}\left(\begin{array}{cc} 1 & 1-2p\\ 1-2p& 1\end{array}\right)$ & $(1-p)^*$ & $p$ &  $p$\\
		\hline
		$\frac{3\pi}{2}$ &$\frac{\pi}{2}$ & $1-2p$ &$\frac{1}{2}\left(\begin{array}{cc} 1 & 1-2p\\ 1-2p& 1\end{array}\right)$ & $p$ & $(1-p)^*$ &  $p$ \\
		\hline
		$\frac{3\pi}{2}$ &$\frac{3\pi}{2}$ & $ 2p-1$ &$\frac{1}{2}\left(\begin{array}{cc} 1 & 2p-1\\ 2p-1& 1\end{array}\right)$ & $(1-p)^*$ & $p$ &  $p$\\
		\hline\hline
	
	\end{tabular}
	\caption{Different final states of the sequential QSS protocol (figure \ref{QSSFigureLabel}) in the presence of bit-flip noise  and the corresponding probabilities for obtaining them. }
			\label{QT}
\end{table}
From the table we see that in 4 out of  $8$ cases the probability of error is $p$, making the total probability of error equal to $\frac{p}{2}$, in accord with equation (\ref{perrorb}).

\section{Conclusion} \label{sec5}
We have used the vectorized form of quantum channels and quantum operations to study the effect of different gate imperfections and different kinds of noise on sequential quantum secret sharing schemes. The noise model considered are dephasing, depolarizing and bit-flip channels. Quite similar analysis and results can also be obtained for phase-flip channel and the amplitude damping channels, although for the sake of brevity the corresponding analysis has not been reported. We have considered only one type of scheme \cite{Schmidt} based on sequential manipulations of qubits and have found the accumulated error probability cannot render the protocol ineffective.  In all the noise models, we have found that in order to extract an error-free shared secret key, the tolerable noise parameter in the channels and the tolerable error in gates scales linearly with the  error and inversely with the number of players in the scheme. Although we have studied only one type of QSS scheme, the method is fairly general to apply it for other schemes and for other noise models. This means essentially that the QSS scheme of \cite{Schmidt} and hence the other sequential models \cite{KA, Tavak} are effective even in the presence of noise.    An open problem even in the absence of noise is how to perform QSS with (k,n) access structures in a sequential way and without entanglement. These are schemes in which any subset of $k$ players out of $n$ players can retrieve the key and subsets of lower size cannot \cite{Lance, Markham, Lau, WuCaiHeZhang}.

\section{Acknowledgements} 
We would like to thank Abdus Salam International Center for Theoretical Physics (ICTP), where part of this research was carried out. We also thank Fabio Benatti for a useful discussion.

{}


\begin{thebibliography}{}

\bibitem{BB84} C. H. Bennett and G. Brassard. "Quantum cryptography: Public key distribution and coin tossing". In Proceedings of IEEE International Conference on Computers, Systems and Signal Processing, volume 175, page 8. New York, 1984.

\bibitem{Ekert}  A. Ekert,  Physical Review Letters, 67, pp.661-663, (1991).

\bibitem{BHB} M. Hillery, V. Buˆzek and A. Berthiaume Phys. Rev. A 59, 1829 (1999).




\bibitem{SenZuk}[8] A. Sen De, U. Sen and M. Zukowski, Phys. Rev. A 68, 032309 (2003).

\bibitem{Pan} Li Xiao, Gui Lu Long, Fu-Guo Deng, and  Jian-Wei Pan, Phys. Rev. A 59, 1829 (1999). 


\bibitem{Karlsson} A. Karlsson, M. Koashi, and N. Imoto,  Phys. Rev. A 59, 162
(1999).

\bibitem{xiao} L. Xiao, G. Lu Long, F.-G. Deng, and J.-W. Pan,  Phys. Rev. A
69, 052307 (2004).

\bibitem{zhang} Z.-j. Zhang and Z.-x. Man, Phys. Rev. A 72, 022303 (2005).

\bibitem{deng} F.-G. Deng, G. L. Long, and H.-Y. Zhou, Phys. Lett. A 340, 43
(2005).






\bibitem{BaghKari} S. Bagherinezhad and V. Karimipour, Phys. Rev. A 67, 044302
(2003).
                                       





\bibitem{Chen} Y.-A. Chen et al., Phys. Rev. Lett. 95, 200502 (2005). 

\bibitem{Gaert} S. Gaertner, C. Kurtsiefer, M. Bourennane and H. Weinfurter Phys. Rev. Lett. 98, 020503 (2007).

\bibitem{TycSanders} Tom\'{a}\v{s} Tyc and Barry C. Sanders, Phys. Rev. A 65, 042310 (2002).

\bibitem{Lance1} Andrew M. Lance, Thomas Symul, Warwick P. Bowen, Tom\'{a}\v{s} Tyc,   Barry C. Sanders, and Ping Koy Lam, New J. Phys. 5 (2003) 4.

\bibitem{Lance2}  Andrew M. Lance, Thomas Symul, Warwick P. Bowen, Barry C. Sanders, and Ping Koy Lam, Phys. Rev. Lett. 92, 177903 (2004) 

\bibitem{GrosshansNature} Fr\ ́{e}d\`{e}ric Grosshans,  Gilles Van Assche,  J\ ́{e}\^{r}ome Wenger,
Rosa Brouri,  Nicolas J. Cerf,  and Philippe Grangiera, Nature 421, 238-241 (2003). 

\bibitem{GrosshansPRL} Fr\`{e}d\`{e}ric Grosshans and Philippe Grangier, Phys. Rev. Lett. 88, 057902 (2002).

\bibitem{Schmidt} 
C. Schmid, P. Trojek, M. Bourennane, C. Kurtsiefer, M. Zukowski, and H. Weinfurter,
Phys. Rev. Lett. 95, 230505 (2005). 

\bibitem{YanGao} Feng-Li Yan, and  Ting Gao, Physical Review A 72, 012304 (2005).



\bibitem{He} G.-P. He Phys. Rev. Lett. 98, 028901 (2007); for the reply
see C. Schmid, P. Trojek,P, M. Bourennane, C. Kurtsiefer, M. Zukowski, and H. Weinfurter,
Rev. Lett. 98, 028902 (2007).


\bibitem{KA}
Vahid Karimipour, Marzieh Asoudeh,
 Phys. Rev. A 92, Rapid Communications, 030301 (2015).

\bibitem{Tavak} A. Tavakoli, I. Herbauts, M.  Zukowski, and  M.  Bourennane, preprint, arXiv:1501.05582.

   \bibitem{chen} K. Chen and H.-K. Lo, Quant. Inf. Comput. 7, 689 (2007).

\bibitem{adesso} 
Ioannis Kogias, Yu Xiang, Qiongyi He, Gerardo Adesso, Phys. Rev. A 95, 012315 (2017).
\bibitem{chat} Maharshi Ray, Sourav Chatterjee, Indranil Chakrabarty,
 Phys. J. D (2016) 70: 114.
\bibitem{agrawal}
Satyabrata Adhikari, Indranil Chakrabarty, Pankaj Agrawal,
Quantum Information and Computation, 12, 0253 (2012).
\bibitem{bohmann} M. Bohmann, J. Sperling, W. Vogel
 Phys. Rev. A 91, 042332 (2015);  M. Bohmann, J. Sperling, W. Vogel, Phys. Rev. A 96, 012321 (2017).






\bibitem{Lance} A. M. Lance, T. Symul, W. P. Bowen, T. Tyc, B. C. Sanders,
and P. K. Lam, New J. Phys. 5, 4 (2003).
\bibitem{Markham} D. Markham and B. C. Sanders, Phys. Rev. A 78, 042309,(2008).

\bibitem{Lau} H.-K. Lau and C. Weedbrook, Phys. Rev. A 88, 042313 (2013). 

\bibitem{WuCaiHeZhang} Y. Wu, R. Cai, G. He, and J. Zhang, Quant. Inf. Proc. 13, 1085
(2014).




\end{thebibliography}
\end{document}